\documentclass[pre,showpacs,showkeys,twocolumn,floatfix,amsmath,amsfonts]{revtex4}

\usepackage{graphicx}
\usepackage{epstopdf}
\usepackage{epsfig}
\usepackage{textcomp}
\usepackage[english]{babel}
\usepackage[usenames, dvipsnames]{color}
\usepackage{color}
\usepackage{soul}

\usepackage[colorlinks]{hyperref}

\hypersetup{colorlinks=true,citecolor=blue,linkcolor=blue,urlcolor=blue}

\begin{document}

\title{Rotobreathers in a chain of coupled elastic rotators}

\author{Alexander~V.~Savin$^{1,2}$}
\email{asavin00@gmail.com}
\author{Ilsiya~R.~Sunagatova$^{3,4}$}
\email{sunagatova66@gmail.com}
\author{Sergey~V.~Dmitriev$^{4,5}$}
\email{dmitriev.sergey.v@gmail.com}

\affiliation{
$^1$N. N. Semenov Federal Research Center for Chemical Physics of RAS (FRCCP RAS), Moscow 119991, Russia \\
$^2$Plekhanov Russian University of Economics, Moscow 117997, Russia\\
$^3$Bashkir State University, Zaki Validi Str. 32, Ufa 450076, Russia\\
$^4$Institute of Molecule and Crystal Physics, Ufa Federal Research Centre of RAS, Oktyabrya Ave. 71, Ufa 450054, Russia\\
$^5$Institute of Mathematics with Computing Centre, Ufa Federal Research Centre of RAS, Chernyshevsky St. 112, Ufa 450008, Russia
}


\begin{abstract}
Rotobreathers in the chain of coupled linearly elastic rotators are analyzed. Each rotator is a particle connected by a massless elastic rod with a frictionless pivot; it has two degrees of freedom, length and angle of rotation. The rods of the rotators and the elastic bonds between the nearest rotators are linearly elastic, and the nonlinearity of the system is of a purely geometric nature. It is shown that long-lived rotobreathers can exist if the stiffness of the rods is high enough to create a relatively wide gap in the phonon spectrum of the chain. The frequency of angular rotation of the rotobreather cannot be above the optical band of the phonon spectrum and is in the spectrum gap. Generally speaking, the rotation of the rotobreather is accompanied by radial oscillations, however, one can choose such initial conditions so that the radial oscillations are minimal. Some parameters of rotobreathers with minimal radial vibrations are presented on the basis of numerical simulations. The results obtained qualitatively describe the behavior of physical systems with coupled rotators.
\end{abstract}

\maketitle
        

\section{Introduction}
\label{Introduction}

Mechanisms of energy localization and transport in nonlinear discrete systems are attracting a lot of attention from physicists because they play a decisive role in a variety of processes. Topological solitons~\cite{BK,ChaosWe}, shock waves~\cite{Shock1,Shock2,Shock3}, crowdions~\cite{Crowd1,Crowd2,Crowd3}, discrete breathers~\cite{Flach1998,Flach2008,UFN,DB,DB1}, modes localized on defects~\cite{DefectMode}, rotobreathers~\cite{Takeno140,Aubry201} are examples of spatially localized objects that exist in nonlinear lattices. 

Dynamics of coupled rotators has been analyzed in the early works by Benettin {\it et al.}~\cite{Benettin89,Benettin103} and later in the works~\cite{Osipov1,Osipov2}.

Rotobreathers are observed experimentally in superconducting Josephson junction arrays~\cite{Ustinov,Ustinov016603,Takeno213,Mazo733,Machida024523,JJ,JJ1,JJ2} and in a polymer crystal which consists of 1D columns of nested rotors arranged in helical arrays~\cite{Monkey}. The single-crystal neutron-diffraction technique was used to analyze the structure of the 4-methylpyridine crystal with the methyl groups rotating about the c axis, revealing the breather modes~\cite{Fillaux}.

The rotor lattice model~\cite{Benettin89,Livi539,LiviNew,Iacobucci2021} was used in the work~\cite{Xiong012125} to show that strength of the thermal rectification effect can increase in the thermodynamic limit in contrast to earlier work on the Frenkel-Kontorova model~\cite{19}. The underlying mechanism is transition from anomalous to normal heat conduction with increasing temperature~\cite{25,26,Savin355}, which is due to the excitation of rotobreathers at high temperatures. The model of coupled rotators~\cite{Benettin89} found its application in describing the relative rotation of polymer fragments around the axis of the macromolecule~\cite{Gendelman591} and it was shown that a strong effect of thermal rectification is possible in a system of polyethylene nanofibers~\cite{8}, possibly due to the excitation of rotobreather modes~\cite{Xiong012125}. Appearance of chaos and synchronization structures in the chains of rotating pendulums have been analyzed in the works~\cite{Bolotov1,Bolotov2,Bolotov3,Bolotov4}.

\textcolor{black}{Introducing additional degrees of freedom into nonlinear chains helps capture some of the new physical effects~\cite{Kofane,Kofane1}.} 

Recently, rotational dynamics of molecules was studied in molecular crystals such as fullerites~\cite{Bub1,Bub2}, chain (or column) of disc-shaped B$_{42}$ molecules~\cite{Bub3} and carbon nanotube bundles~\cite{JMMP}. Rotobreathers can be excited thermally~\cite{Takeno140} and hence they can contribute to heat capacity of the molecular crystals.

\textcolor{black}{When considering complex nonlinear lattices, such as molecular crystals with many degrees of freedom per rotating particle, rigorous proof of the existence of rotobreathers as time-periodic dynamic regimes becomes problematic. Numerical analysis of real crystals is always based on a number of approximations, for example, on the use of phenomenological interatomic potentials, and even the mass of an atom is a probabilistic characteristic due to the presence of isotopes. The chain of rotators considered here is not as complicated as real crystals, but, nevertheless, the problem of finding exact solutions is deliberately replaced by the search for long-lived rotobreathers, which can be obtained using very simple initial conditions. Finding exact solutions remains an important issue and must be done in future works. For discrete breathers, a step towards real lattices was made in~\cite{Quasi}, where the concept of quasi-breathers was proposed.} 

Most of the analyzed chains supporting rotobreathers had one rotational degree of freedom per particle~\cite{25,26,Savin355,R1,R2,R3,R4,Rotobreathers}, \textcolor{black}{although Josephson junctions are described by models with two degrees of freedom per site~\cite{JJ}}. In a chain of connected beads sliding along rigid rings considered in~\cite{Rotobreathers}, it was shown that rotobreathers have no upper limit on the rotation frequency. 

Here we consider a chain of coupled elastic rotators with two degrees of freedom, radial and angular, and demonstrate that, due to the finite rigidity of the linearly elastic rotators, the angular rotation frequency of rotobreathers cannot exceed the optical band of the phonon spectrum. 

The chain of coupled rotators is described in Sec.~\ref{ChainModel}, phonon spectra for the chain in the ground states are analyzed in Sec.~\ref{Disp}, rotobreathers are modelled in Sec.~\ref{Rotobreathers}, and conclusions are drown in Sec.~\ref{Conclusions}.
\begin{figure}[tb]
\begin{center}
\includegraphics[angle=0, width=1.0\linewidth]{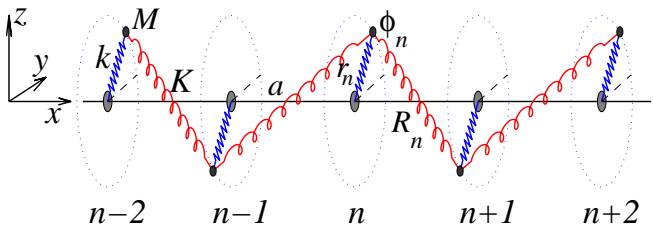}
\end{center}
\caption{\label{Fig1}\protect
Schematic of a chain of coupled elastic rotators numbered by the index $n$. A rotator is a point-wise particle of mass $M$ connected by a massless, linear elastic rod of stiffness $k$ to a frictionless pivot. Particles rotate parallel to the $(y,z)$ plane about common rigid spoke, which is parallel to the $x$ axis. Each particle is coupled to the nearest neighbours by linear elastic bonds of stiffness $K$. The rods and bonds have equilibrium lengths $r_0$ and $R_0$, respectively, and current lengths $r_n$ and $R_n$, respectively.
}
\end{figure}

\section{Chain of coupled elastic rotators}
\label{ChainModel}

Consider a chain of coupled elastic rotators numbered by the index $n$ and spaced apart by a distance $a$, as shown in Fig.~\ref{Fig1}. Each rotator is a point-wise particle of mass $M$ connected by a massless, linear elastic rod of stiffness $k$ to a frictionless pivot. Particles rotate parallel to the $(y,z)$ plane about common rigid spoke, which is parallel to the $x$ axis. Each particle is coupled to the nearest neighbours by linear elastic bonds of stiffness $K$. The rods and bonds have equilibrium lengths $r_0$ and $R_0$, respectively. Each particle has two degrees of freedom, the distance from the spoke, $r_n$, and the angle of rotation, $\phi_n$, counted counterclockwise from the $y$ axis. Coordinates of the $n$-th particle are $(x_n,y_n,z_n)=(na,r_n\cos\phi_n,r_n\sin\phi_n)$ and the distance between particles $n$ and $n+1$ is
\begin{eqnarray}
R_n=\sqrt{a^2+r_{n+1}^2+r_n^2-2r_nr_{n+1}\cos(\phi_{n+1}-\phi_n)}.
\label{Rn}
\end{eqnarray}

The Hamiltonian of the chain of rotators has the form
\begin{eqnarray}
H=\sum_n \Big[ \frac M2(r_n^2\dot{\phi}_n^2+\dot{r}_n^2)+\frac k2 (r_n-r_0)^2\nonumber\\
+\frac K2 (R_n-R_0)^2 \Big], \label{f1}
\end{eqnarray}
where overdor means differentiation with respect to time. The first, second, and third terms in the square brackets present the kinetic energy, potential energy of the elastic rods, and potential energy of the elastic bonds, respectively. 
\begin{figure}[tb]
\begin{center}
\includegraphics[angle=0, width=1.\linewidth]{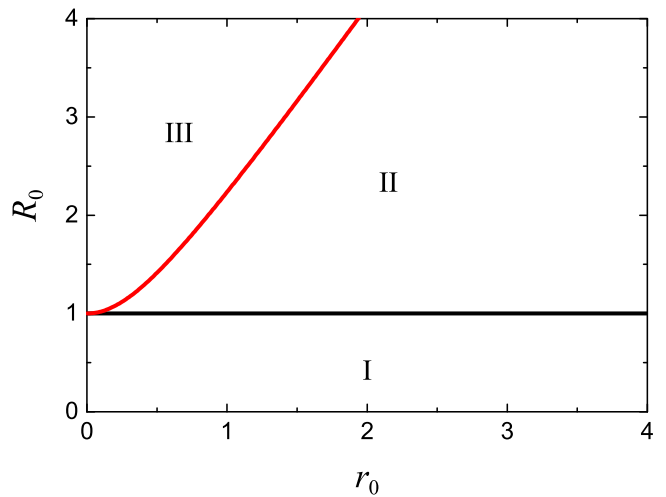}
\end{center}
\caption{Phase diagram of the chain of rotators. Regime I is realized for $R_0\le a=1$, regime III for $R_0\ge \sqrt{a^2+4r_0^2}=\sqrt{1+4r_0^2}$, and regime II in between. Ground states in these regimes are described in the text.
}\label{Fig2}
\end{figure}

With the help of the Hamilton's equation, the following equations of motion can be derived form the Hamiltonian Eq.~(\ref{f1}),
\begin{eqnarray}
M\ddot{r}_n=Mr_n\dot{\phi}_n^2-k(r_n-r_0) \nonumber \\
-K\left(1-\frac{R_0}{R_n}\right)[r_n-r_{n+1} \cos(\phi_{n+1}-\phi_n)] \nonumber \\
-K\left(1-\frac{R_0}{R_{n-1}}\right)[r_n-r_{n-1}\cos(\phi_{n}-\phi_{n-1})], 
\label{EqMor}
\end{eqnarray}
\begin{eqnarray}
M\ddot{\phi}_n=-2M\frac{\dot{r}_n}{r_n}\dot{\phi}_n \nonumber \\
+ K\left(1-\frac{R_0}{R_n}\right) \frac{r_{n+1}}{r_n} \sin(\phi_{n+1}-\phi_n) \nonumber \\
- K\left(1-\frac{R_0}{R_{n-1}}\right) \frac{r_{n-1}}{r_n} \sin(\phi_{n}-\phi_{n-1}). \label{EqMoPhi}
\end{eqnarray}

Out of the six model parameters (particle mass $M$, spring constants $k$ and $K$, distance between rotators $a$, rotator equilibrium length $r_0$, and bond equilibrium length $R_0$) three can be scaled out by proper choice of the units of time, distance, and energy. With this in mind, in the numerical examples we will always set $M=1$, $a=1$, and $K=1$ and study the effect of the remaining parameters, $r_0$, $R_0$, and $k$.

Similar to the beads and rings model~\cite{Rotobreathers}, considered chain of rotators supports three different ground state structures depending on the geometry parameters $r_0$ and $R_0$, as shown in the phase diagram, Fig.~\ref{Fig2}. In regime I, which is realized for $R_0\le a=1$, all rotators have equilibrium length $r_n=r_0$, all bonds are extended, $R_n=a>R_0$, and $\phi_n=\phi={\rm const}$. Regime III is observed for $R_0\ge \sqrt{a^2+4r_0^2}=\sqrt{1+4r_0^2}$. In this regime, rotators are extended, $r_n>r_0$, bonds are compressed $R_n<R_0$, and $\phi_{n+1}-\phi_n=\pi$. Regime II is realised for the portion of the phase diagram in between regimes I and III. In this regime, $r_n=r_0$, $R_n=R_0$ and, as follows from Eq.~(\ref{Rn}),
\begin{equation}
\cos(\phi_{n+1}-\phi_n)=1+\frac{a^2-R_0^2}{2r_0^2}. \label{XXf1}
\end{equation}

In regime II, the structure of the chain is indefinite because the sign of $ \phi_{n+1}-\phi_n \equiv \Delta\phi_n$ in Eq.~(\ref{XXf1}) can be arbitrary. The structure, for example, can be chiral if all $\Delta\phi_n$ are of the same sign, it can have a zigzag structure with alternating signs of $\Delta\phi_n$, or it can be random if the sign of $\Delta\phi_n$ is chosen randomly.

In the subsections \ref{RegimeIrot} and \ref{RegimeIIIrot}, rotobreathers will be analyzed in well-defined structures I and III, respectively.
\textcolor{black}{We will take $r_0=0.5$ and two values of the parameter $R_0$, namely 0.8 and $2\sqrt{2}=2.828$, at which regimes I and III, respectively, are realized relatively far from the borders of their existence.}

\section{Dispersion relations for ground states}
\label{Disp}

Spatially localized dynamic modes, including rotobreathers, should have frequencies outside the phonon spectrum. Therefore, it is important to obtain dispersion relations for low-amplitude oscillations around the ground states of regimes I and III.

\subsection{Regime I}
\label{RegimeI}

In the case
\begin{equation}
R_0\le a, 
\label{ConditionI}
\end{equation}
the ground state of the considered system is %
\begin{equation}
r_n=r_0, \quad R_n=a, \quad \phi_n= \phi ={\rm const}. 
\label{groundI}
\end{equation}

Let us consider small perturbation of the ground state
\begin{equation}
r_{n}(t)=r_0+\delta_{n}(t), \quad \phi_{n}(t)=\phi + \epsilon_{n}(t), 
\label{PerturbI}
\end{equation}
where $\delta_{n}(t)\ll r_0$ and $\epsilon_{n}(t)\ll 1$.

Substituting Eq.~(\ref{PerturbI}) into Eqs.~(\ref{EqMor}), (\ref{EqMoPhi}) and keeping in the Taylor series expansions only up to linear terms in $\delta_{n}$ and $\epsilon_{n}$ one obtains the following linearized equations of motion
\begin{eqnarray}
M\ddot{\delta}_n=-k\delta_n \nonumber \\
+K\left(1-\frac{R_0}{a} \right) (\delta_{n-1}-2\delta_n+\delta_{n+1}), 
\label{f9LinI}
\end{eqnarray}
\begin{eqnarray}
M\ddot{\epsilon}_n=K\left(1-\frac{R_0}{a}\right) (\epsilon_{n-1}-2\epsilon_n+\epsilon_{n+1}).
\label{f10LinI}
\end{eqnarray}
It can be seen that the linearized equations of motion are decoupled and rotational and radial displacements become independent.
\begin{figure}[tb]
\begin{center}
\includegraphics[angle=0, width=1.0\linewidth]{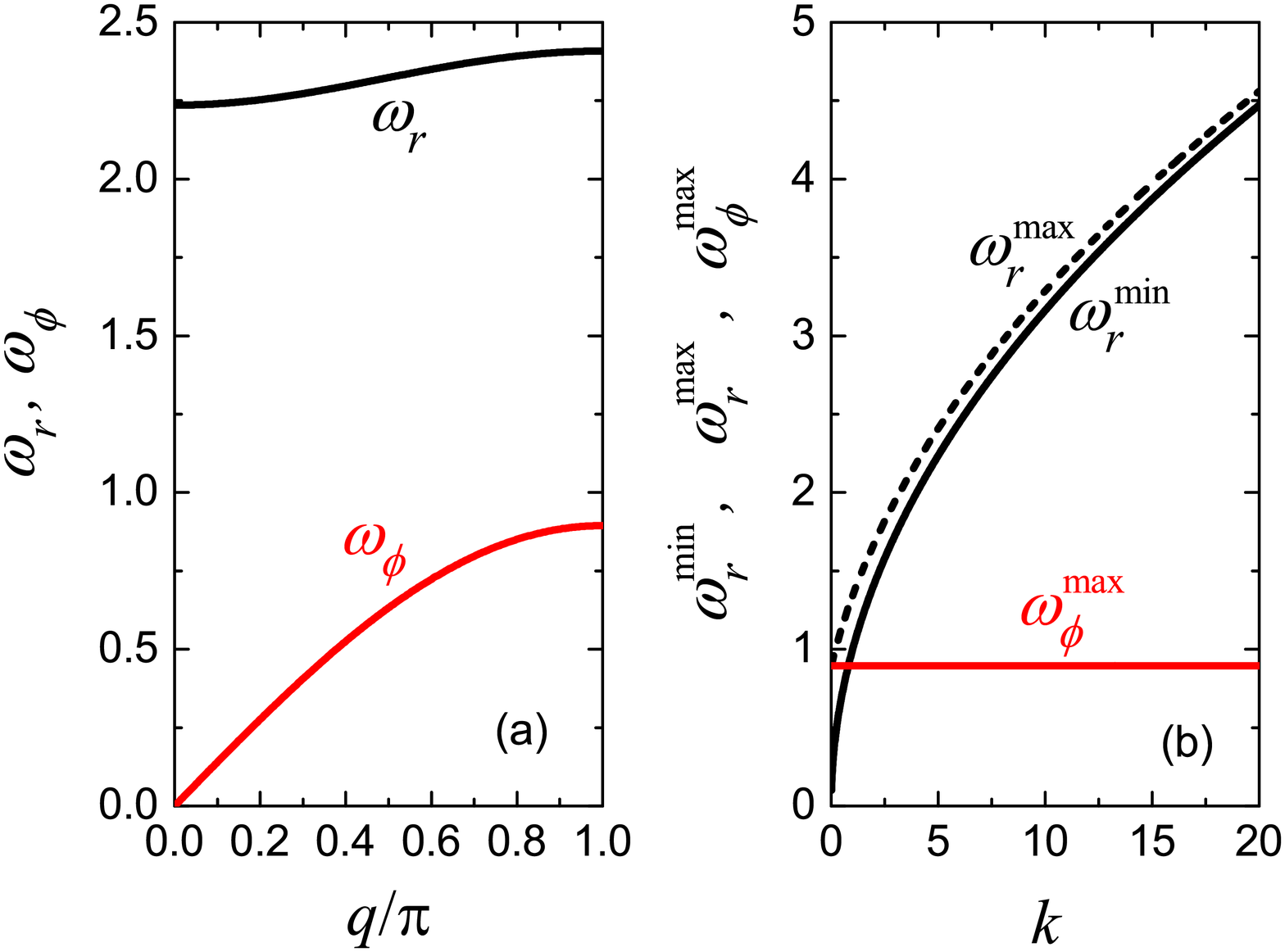}
\end{center}
\caption{(a) Example of phonon dispersion curves for the ground state in regime I, Eqs.~(\ref{DispDelI}) and (\ref{DispEpsI}), for $r_0=0.5$, $R_0=0.8$, and $k=5$. (b) Maximal acoustic frequency, $\omega_{\phi}^{\rm max}$, and minimal and maximal optic frequencies, $\omega_{r}^{\rm min}$ and $\omega_{r}^{\rm max}$, as the functions of $k$.
}\label{Fig3}
\end{figure}

Looking for the solution of the equations of motion Eqs.~(\ref{f9LinI}) and (\ref{f10LinI}) in the form $\delta_n\sim \exp[i(qn-\omega_r t)]$ and $\epsilon_n\sim \exp[i(qn-\omega_{\phi} t)]$ one comes to the dispersion relations
\begin{eqnarray}
\omega_r=\sqrt{ \frac{k}{M} +\frac{4K}{M} \left(1-\frac{R_0}{a} \right)\sin^2 \frac{q}{2} },
\label{DispDelI}
\end{eqnarray}
\begin{eqnarray}
\omega_{\phi}=2\sqrt{\frac{K}{M}\left(1-\frac{R_0}{a}\right)}\sin\frac{q}{2}.
\label{DispEpsI}
\end{eqnarray}

Example of phonon dispersion curves, Eqs.~(\ref{DispDelI}) and (\ref{DispEpsI}), is given in Fig.~\ref{Fig3}(a) for $r_0=0.5$, $R_0=0.8$, and $k=5$. Optic and acoustic bands are presented by $\omega_{r}$ and $\omega_{\phi}$, respectively. In Fig.~\ref{Fig3}(b), as the functions of $k$, maximal acoustic frequency, $\omega_{\phi}^{\rm max}$, and minimal and maximal optic frequencies, $\omega_{r}^{\rm min}$ and $\omega_{r}^{\rm max}$, are plotted. It can be seen from the figure and from Eq.~(\ref{DispEpsI}) that the acoustic phonon frequencies are $k$-independent. On the other hand, optic frequencies increase with increasing $k$. 
\begin{figure}[tb]
\begin{center}
\includegraphics[angle=0, width=1.0\linewidth]{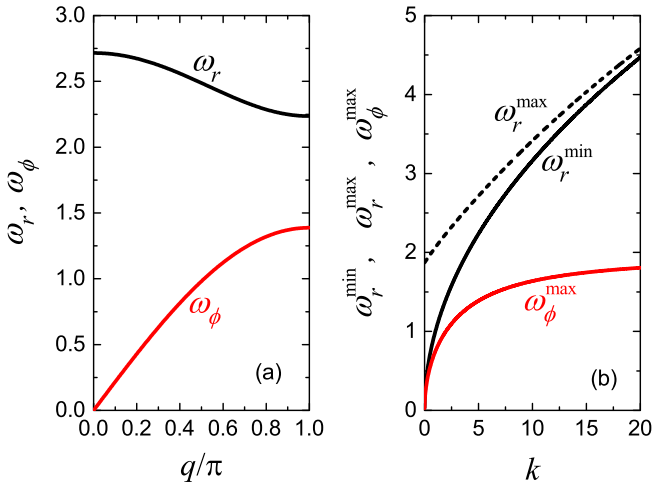}
\end{center}
\caption{(a) Example of phonon dispersion curves for the ground state in regime III, Eqs.~(\ref{DispDel}) and (\ref{DispEps}), for $r_0=0.5$, $R_0=2\sqrt 2$, and $k=5$. (b) Maximal acoustic frequency, $\omega_{\phi}^{\rm max}$, and minimal and maximal optic frequencies, $\omega_{r}^{\rm min}$ and $\omega_{r}^{\rm max}$, as the functions of $k$.
}\label{Fig4}
\end{figure}

\subsection{Regime III}
\label{RegimeIII}

In the case 
\begin{equation}
R_0\ge \sqrt{a^2+4r_0^2}, 
\label{ConditionIII}
\end{equation}
the ground state is 
\begin{equation}
r_n=r={\rm const}, \quad \phi_n=(-1)^n\frac{\pi}{2} + \phi, 
\label{groundIII}
\end{equation}
with arbitrary constant $\phi$ and
$r$ found as a minimum of the potential energy per atom
\begin{equation}
P(r)=\frac{k}{2}(r-r_0)^2+\frac{K}{2}\left(\sqrt{a^2+4r^2}-R_0 \right)^2.
\label{f3}
\end{equation}
Condition for the minimum of the function $P(r)$, $dP/dr=0$, leads to the algebraic equation of the fourth order which is solved numerically by the Newton-Raphson method.

Let us consider small perturbation of the ground state
\begin{equation}
r_{n}(t)=r+\delta_{n}(t), \quad \phi_{n}(t)=(-1)^n\frac{\pi}{2}+\epsilon_{n}(t), 
\label{Perturb}
\end{equation}
where $\delta_{n}(t)\ll r$ and $\epsilon_{n}(t)\ll 1$. 

Substituting Eq.~(\ref{Perturb}) into Eqs.~(\ref{EqMor}), (\ref{EqMoPhi}) one can obtain the following linearized equations of motion
\begin{eqnarray}
M\ddot{\delta}_n=-k\delta_n \nonumber \\
-K\left(1-\frac{R_0}{L} +\frac{4R_0r^2}{L^3}\right)(\delta_{n-1}+2\delta_n+\delta_{n+1}), 
\label{f9Lin}
\end{eqnarray}
\begin{eqnarray}
M\ddot{\epsilon}_n=-K\left(1-\frac{R_0}{L}\right) (\epsilon_{n-1}-2\epsilon_n+\epsilon_{n+1}),
\label{f10Lin}
\end{eqnarray}
where
\begin{equation}
L=\sqrt{a^2+4r^2}.
\label{L0}
\end{equation}
In this case, the linearized equations of motion are also decoupled.

The dispersion relations for the radial and rotational displacements are found by substituting the ansatz $\delta_n\sim \exp[i(qn-\omega_r t)]$ into Eq.~(\ref{f9Lin}) and $\epsilon_n\sim \exp[i(qn-\omega_{\phi} t)]$ into Eq.~(\ref{f10Lin}). The result reads
\begin{eqnarray}
\omega_r=\sqrt{ \frac{k}{M} +\frac{4K}{M} \left(1-\frac{R_0}{L} +\frac{4R_0r^2}{L^3}\right) \left( 1 - \sin^2 \frac{q}{2} \right)},
\label{DispDel}
\end{eqnarray}
\begin{eqnarray}
\omega_{\phi}=2\sqrt{\frac{K}{M}\left(\frac{R_0}{L}-1\right)}\sin\frac{q}{2}.
\label{DispEps}
\end{eqnarray}

The dependencies Eqs.~(\ref{DispDel}) and (\ref{DispEps}) are presented in Fig.~\ref{Fig4}(a) for $r_0=0.5$, $R_0=2\sqrt{2}$, and $k=5$. In Fig.~\ref{Fig4}(b), maximal acoustic frequency, $\omega_{\phi}^{\rm max}$, and minimal and maximal optic frequencies, $\omega_{r}^{\rm min}$ and $\omega_{r}^{\rm max}$, are plotted as the functions of $k$. In this case, the acoustic frequencies $\omega_{\phi}$ depend on $k$ through $L$ given by Eq.~(\ref{L0}), because equilibrium length of rotators $r$ corresponds to the minimum of function Eq.~(\ref{f3}), which includes $k$. Optic frequencies $\omega_r$ increase with increasing $k$.

\section{Rotobreathers}
\label{Rotobreathers}

It will be shown that the rotobreather frequency cannot be higher than the optical band of the phonon spectrum, that is, its frequency must be in the gap between the optical and acoustic bands. It is clear that rotobreathers cannot exist in a chain of rotators with very small $k$, since in this case the gap is either absent [see Fig.~\ref{Fig3}(b) for regime I] or is too narrow, while the second harmonic lies in the optic band [Fig.~\ref{Fig4}(b) for regime III]. Therefore, we will consider chains with sufficiently wide gaps in the phonon spectrum.

First, the anti-continuum limit with non-interacting rotators will be considered, and then rotobreathers in the chain of rotators will be analysed in regimes I and III.

\subsection{Single elastic rotator}
\label{ElasticRotator}

Let us consider the anti-continuum limit by setting $K=0$; in this case the rotators become uncoupled. The Hamiltonian Eq.~(\ref{f1}) for single rotator simplifies to
\begin{equation}
H= \frac M2(r^2\dot{\phi}^2+\dot{r}^2)+\frac k2 (r-r_0)^2, \label{RotatorHam}
\end{equation}
where the first and the second terms in the right-hand side give the kinetic and potential energies of the rotator, respectively.

The equations of motion Eqs.~(\ref{EqMor}) and (\ref{EqMoPhi}) obtain the form
\begin{eqnarray}
M\ddot{r}&=&Mr\dot{\phi}^2-k(r-r_0), \label{EqMor1} \\
\ddot{\phi}&=&-2\frac{\dot{r}}{r}\dot{\phi}. \label{EqMoPhi1}
\end{eqnarray}

In the absence of rotation, i.e., for $\phi(t)={\rm const}$, and hence $\dot{\phi}(t)=0$, Eq.~(\ref{EqMor1}) describes harmonic oscillations of the rotator radius with frequency
\begin{equation}
\Omega_r= \sqrt{\frac{k}{M}}. \label{Omr}
\end{equation}

Next, consider the vibration-free rotation of the rotator for which
\begin{equation}
r(t)= r= {\rm const}. \label{rconst}
\end{equation}
Then from Eq.~(\ref{EqMoPhi1}) one has $\ddot{\phi}=0$ and hence, $\dot{\phi}= {\rm const}$. Moreover, from Eq.~(\ref{EqMor1}) it follows that
\begin{equation}
\dot{\phi}_v=\sqrt{\frac{k(r-r_0)}{Mr}}. \label{dotphi}
\end{equation}

Period of rotation and angular frequency of the vibrationless rotobreather are
\begin{equation}
T=\frac{2\pi}{\dot{\phi}_v}, \quad \Omega_{\phi}=\frac{2\pi}{T}=\dot{\phi}_v, \label{TOmPhi}
\end{equation}
respectively

Energy of the vibrationless rotobreather can be obtained by substituting $\dot{r}=0$ and Eq.~(\ref{dotphi}) into Eq.~(\ref{RotatorHam}). The result reads
\begin{equation}
H= \frac {kr}{2}(r-r_0)+\frac k2 (r-r_0)^2. \label{HamVL}
\end{equation}

Equation~(\ref{HamVL}) shows that with increasing $r$, the total energy of the vibrationless rotobreather diverges as $\sim r^2$. Kinetic energy of rotobreather is greater than the potential energy and the difference between them vanishes in the limit $r \rightarrow \infty$, when the circular motion becomes rectilinear. 

Interestingly, from Eqs.~(\ref{dotphi}) and (\ref{TOmPhi}) it follows that, with increasing total energy of the rotator, in the limit $r \rightarrow \infty$, the angular frequency of the rotobreather approaches the value 
\begin{equation}
\Omega_{\phi} \rightarrow \sqrt{\frac kM} =\Omega_{r}. \label{Tw}
\end{equation} 

We conclude that the angular frequency of rotations $\Omega_{\phi}$ increases with $r$ (i.e., it increases with total energy of the rotobreather) but it cannot exceed the frequency of radial vibration $\Omega_r$. It will be shown that frequency of rotobreathers in the chain of rotators also cannot exceed frequency of radial vibrations.

\subsection{Regime I}
\label{RegimeIrot}

\textcolor{black}{Note that this work does not pose the problem of finding rotobreathers that are strictly periodic in time; therefore, the simplest initial conditions are used, when at $t=0$ one rotator is excited in the middle of the chain, while the other rotators are initially in their equilibrium positions. Absorbing boundary conditions are used to exclude the influence of the radiation of the central rotator on its dynamics. The typical number of rotators in a chain is $N=300$, with 100 rotators in the middle without attenuation, and 100 rotators in the left and right parts of the chain are used to absorb radiation. This size of the computational cell was sufficient, since only sharply localized rotobreathers were analyzed.}

The ground state of the chain of coupled rotators $(K>0)$ in regime I is described by Eq.~(\ref{groundI}). The following initial conditions are used. For the rotator in the middle of the chain, $n=N/2$, we set
\begin{eqnarray}
r_{N/2}(0)=r^*, \quad \dot{r}_{N/2}(0)=0, \nonumber\\ \phi_{N/2}(0)=0, \quad
\dot{\phi}_{N/2}(0)=\dot{\phi}^*, 
\label{ICone}
\end{eqnarray}
which means that the rotator at $t=0$ has initial length $r^*$, zero initial radial velocity, zero initial angle and initial angular velocity $\dot{\phi}^*$. All other rotators are in their ground states with zero initial velocities
\begin{eqnarray}
r_n(0)&=&r_0, \quad \dot{r}_n(0)=0, \nonumber\\ \phi_n(0)&=&0, \quad
\dot{\phi}_n(0)=0, \quad n\neq N/2.
\label{ICall}
\end{eqnarray}

Under these initial conditions, the rotobreather kinetic energy at $t=0$ is equal to 
\begin{equation}
T^*=\frac{M}{2}(r^*)^2(\dot{\phi}^*)^2.
\label{Tinit}
\end{equation}
The change in the potential energy at $t=0$ is
\begin{eqnarray}
\Delta P^*=\frac{k}{2}(r^*-r_0)^2 + K(R^*-R_0)^2 
-K(a-R_0)^2,
\label{PinitBad}
\end{eqnarray}
where the initial length of the bonds connecting rotator $n=N/2$ with its neighbors is
\begin{eqnarray}
R^*=\sqrt{a^2+r_{0}^2+(r^*)^2-2r^*r_0}.
\label{Rn00}
\end{eqnarray}
Note that the third term on the right-hand side of Eq.~(\ref{PinitBad}) is introduced to subtract the potential energy of the ground state.

Initial energy given to the chain is
\begin{eqnarray}
H^*=T^*+\Delta P^*.
\label{H0}
\end{eqnarray}
%
\begin{figure}[tb]
\begin{center}
\includegraphics[angle=0, width=1.0\linewidth]{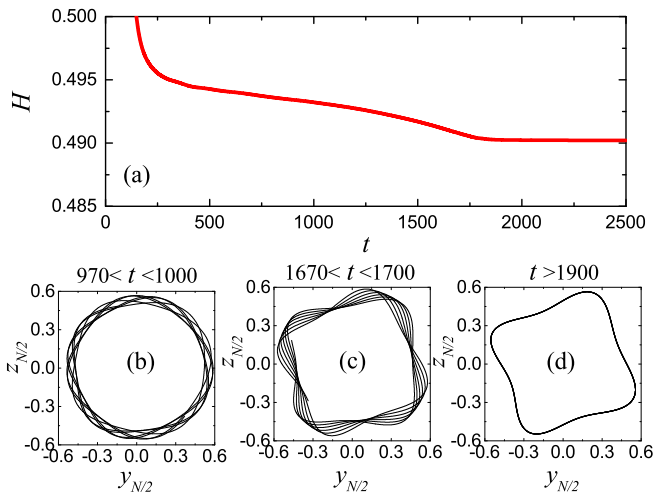}
\end{center}
\caption{\textcolor{black}{Results for regime I. (a) Time evolution of the total energy of the chain with absorbing boundary conditions and central rotator initially excited with $r^*=0.6$ and $\dot{\phi}^*=1.6$. (b-d) Trajectories of the excited rotator on the $(y,z)$ plane for time intervals specified for each case. In (d), a periodic motion of the rotator is observed, since the trajectory is closed. Model parameters: $r_0=0.5$, $R_0=0.8$, and $k=20$.}
}\label{Fig5}
\end{figure}

Let us take the model parameters $r_0=0.5$, $R_0=0.8$, and $k=20$. Dispersion curves for these parameters are presented in Fig.~\ref{Fig3} from which it is seen that for $k=20$ the gap in the phonon spectrum is relatively wide.

\textcolor{black}{Our strategy for searching for quasiperiodic rotobreathers is to excite a central rotator with initial parameters $r^*$ and $\dot{\phi}^*$ and wait until the energy emitted by the rotator is absorbed at the boundaries of the chain. This strategy will produce several families of quasiperiodic rotobreathers, each family is characterised by the topology of the trajectory of the central rotator on the $(y,z)$ plane.}

\textcolor{black}{One example is presented in Fig.~\ref{Fig5} for $r^*=0.6$ and $\dot{\phi}^*=1.6$. In (a), total energy of the chain with absorbing boundary conditions is presented. The total energy decreases over time because the energy emitted by the central rotator is absorbed at the boundaries of the chain. However, the total energy becomes almost constant for $t>1900$, because the emission of energy practically stops. Panels (b-d) show the trajectories of the excited rotator in the $(y,z)$ plane for the time intervals specified for each case. As seen in (d), for $t>1900$, a (quasi)periodic motion of the rotator is observed  because the trajectory is closed.}
\begin{figure}[tb]
\includegraphics[angle=0, width=1.0\linewidth]{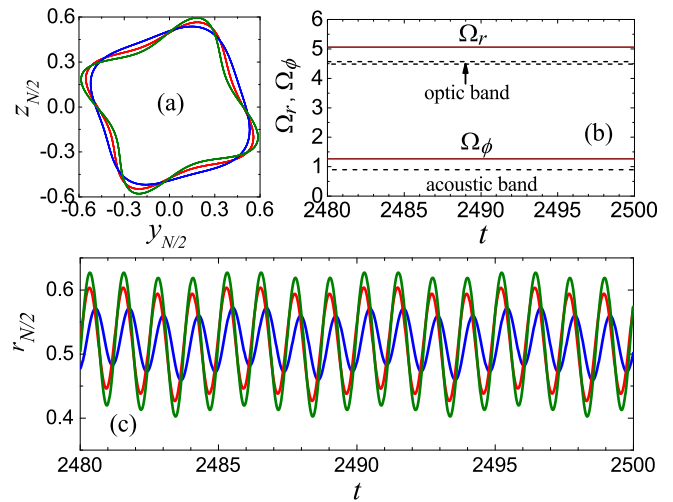}
\caption{\label{Fig6}\protect
\textcolor{black}{Results for regime I. Properties of three rotobreathers obtained with the initial length $r^*=0.6$ and three different values of the initial angular velocity, $\dot{\phi}^*=1.55$, 1.6 and 1.7. In (a), trajectories of the central rotator on the $(y,z)$ plane are shown by the blue, red and green lines, respectively. In (c), the length of the central rotator as the function of time is shown by the blue, red and green lines, respectively. All three rotobreathers have very close angular frequency $\Omega_{\phi}=1.27$ and radial frequency $\Omega_r=4\Omega_{\phi}$, as shown in (b). The dashed lines in (b) show the borders of the phonon spectrum with a very narrow optic band. It can be seen that $\Omega_r$ lies above the optic band and $\Omega_{\phi}$ in the gap of the spectrum. Model parameters: $r_0=0.5$, $R_0=0.8$, and $k=20$.}
}
\end{figure}

\textcolor{black}{In Fig.~\ref{Fig6} we present some characteristics of the quasiperiodic rotobreather shown in Fig.~\ref{Fig5}(d) and two other rotobreathers of this family with similar trajectories on the $(y,z)$ plane. The rotobreathers were obtained with the initial conditions $r^*=0.6$ and three different values of the initial angular velocity, $\dot{\phi}^*=1.55$, 1.6 and 1.7. In Fig.~\ref{Fig6}(a), trajectories of the central rotator on the $(y,z)$ plane are shown by the blue, red and green lines, respectively. In (c), the length of the central rotator as the function of time is shown by the blue, red and green lines, respectively. All three rotobreathers have very close angular frequency $\Omega_{\phi}=1.27$ and radial frequency $\Omega_r=4\Omega_{\phi}$, as shown in (b). Note that the period of the rotational motion is the time required for one complete rotation, while the period of the radial oscillations is calculated as the time between the nearest highs (or lows) of the $r_{N/2}(t)$ curve. Within one angular period of rotation there are four radial oscillation periods, so the trajectory on the $(y,z)$ plane has a squarish shape. The dashed lines in (b) show the borders of the phonon spectrum. It can be seen that $\Omega_r$ lies above the optic band and $\Omega_{\phi}$ in the gap of the spectrum. The absence of resonances with phonons is the reason for the extremely long lifetime of the rotobreathers. It is interesting that a single rotator in the absence of rotation oscillates with an amplitude-independent frequency given by Eq.~(\ref{Omr}), which gives $\Omega_r=4.47$ for the chosen parameters. However in the presence of rotation, due to the geometric nonlinearity, the vibration frequency shifts to the value $\Omega_r=5.08$. Energies of the three obtained rotobreathers are $H=0.446$, 0.490 and 0.538, respectively.}

\textcolor{black}{The value of the initial angular velocity $\dot{\phi}^*$ should be compared to the angular velocity of vibrationless rotation of single rotator, $\dot{\phi}_v$, see Eq.~(\ref{dotphi}). For chosen parameters one has $\dot{\phi}_v=1.83$. The family of rotobreathers shown in Fig.~\ref{Fig6} was excited with the initial angular velocities below $\dot{\phi}_v$; therefore, noticeable radial oscillations can be seen in Fig.~\ref{Fig6}(c). Our next step is to obtain a rotobreather with minimal radial oscillations.} 
\begin{figure}[tb]
\begin{center}
\includegraphics[angle=0, width=1.0\linewidth]{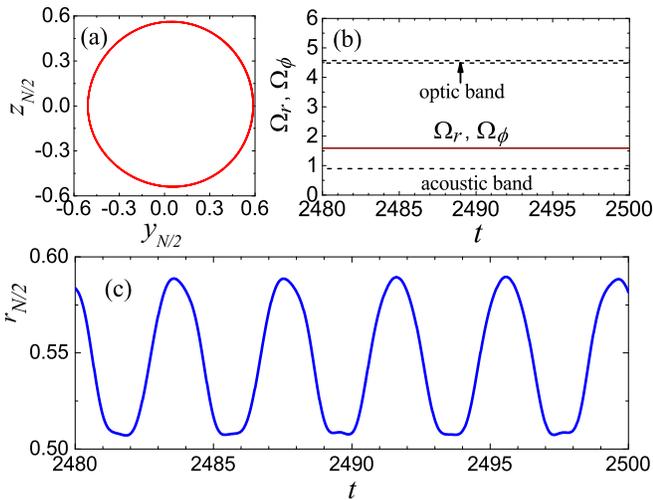}
\end{center}
\caption{\textcolor{black}{Results for regime I. Quasiperiodic rotobreather with minimal radial vibrations obtained with the initial parameters $r^*=0.6$ and $\dot{\phi}^*=1.75$. (a) Nearly circular trajectory of the central rotator on the $(y,z)$ plane. (b) Radial frequency $\Omega_r$ and angular frequency $\Omega_{\phi}$ as the functions of time in the regime of quasiperiodic motion. These frequencies are equal in this case. Dashed lines show the borders of the acoustic and optic bands of the phonon spectrum. (c) Length of the central rotator as the function of time. Model parameters: $r_0=0.5$, $R_0=0.8$, and $k=20$.}
}\label{Fig7}
\end{figure}

\textcolor{black}{For the value $r^*=0.6$, the quasiperiodic rotobreather with minimal radial oscillations is observed for $\dot{\phi}^*=1.75$, which is close to $\dot{\phi}_v=1.83$. Breather practically stops radiating energy at $t=1000$ at the energy level $H=0.607$. Parameters of the rotobreather can be seen in Fig.~\ref{Fig7}. Panel (a) shows the trajectory of the central rotator, which is very close to a circle whose center is offset from the origin in the $y$ direction by 0.040. The most interesting feature of this rotobreather can be seen in (b), that is the equality of the frequencies of radial and rotational motion, $\Omega_r=\Omega_\phi=1.59$. The frequency is in the phonon spectrum gap, and this explains why the rotobreather has an extremely long lifetime. Panel (c) shows the length of the central rotator as the function of time.} 
\begin{figure}[tb]
\includegraphics[angle=0, width=1.0\linewidth]{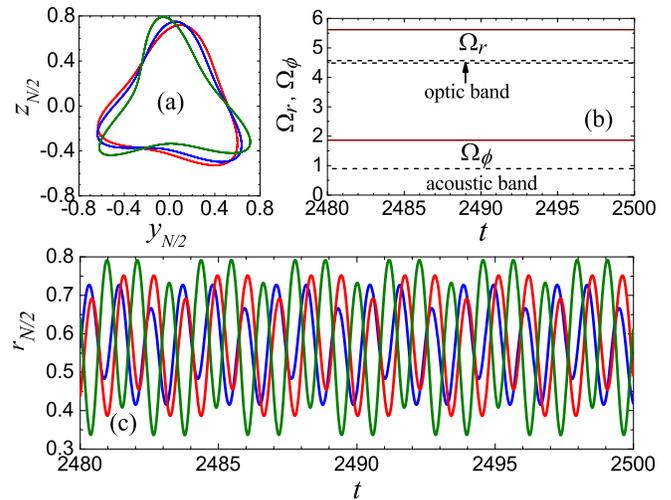}
\caption{\label{Fig8}\protect
\textcolor{black}{Results for regime I. Properties of three quasiperiodic rotobreathers obtained with the initial conditions $r^*=0.6$ and three values of the initial angular velocity, $\dot{\phi}^*=2.3$, 2.4 and 2.6. In (a), trajectories of the central rotator on the $(y,z)$ plane are shown by the blue, red and green lines, respectively. In (c), the length of the central rotator as the function of time is shown by the blue, red and green lines, respectively. All three rotobreathers in the regime of quasiperiodic motion have very close radial frequency $\Omega_r\approx 5.62$ and angular frequency $\Omega_{\phi}\approx 1.86$, as shown in (b). The dashed lines in (b) show the borders of the phonon spectrum. It can be seen that $\Omega_r$ lies above the optic band and $\Omega_{\phi}$ in the gap of the spectrum. Model parameters: $r_0=0.5$, $R_0=0.8$, and $k=20$.}
}
\end{figure}
\begin{figure}[tb]
\includegraphics[angle=0, width=1.0\linewidth]{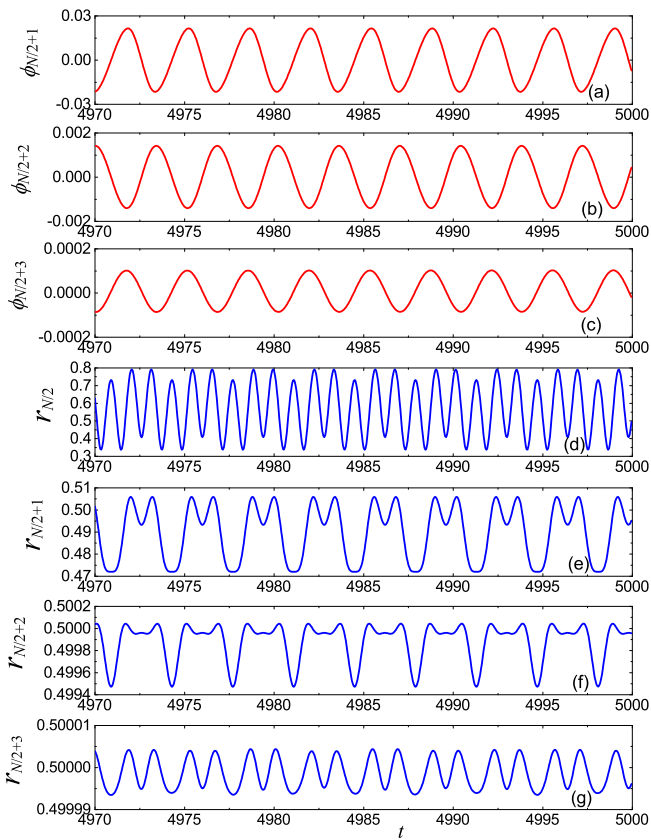}
\caption{\label{Fig8new}\protect
\textcolor{black}{Results for regime I. Time evolution of angular coordinates [panels (a)-(c)] and radial coordinates [panels (d)-(g)] for rotobreather excited with $r^*=0.6$ and $\dot{\phi}^*=2.6$. Model parameters: $r_0=0.5$, $R_0=0.8$, and $k=20$.}
}
\end{figure}

\textcolor{black}{Families of quasiperiodic in time rotobreathers can also be excited with the initial angular velocities above $\dot{\phi}_v$. In Fig.~\ref{Fig8}, a family of rotobreathers obtained with the initial parameters $r^*=0.6$ and $\dot{\phi}^*=2.3$, 2.4, and 2.6 is presented. These initial velocities are noticeably above the velocity of vibrationless rotation of single rotator, $\dot{\phi}_v=1.83$. In (a), closed trajectories of the central rotator on the $(y,z)$ plane are shown by the blue, red and green lines, respectively. In (c), the length of the central rotator as the function of time is shown by the blue, red and green lines, respectively. All three rotobreathers in the regime of quasiperiodic motion have very close angular frequency $\Omega_{\phi}\approx 1.86$ and radial frequency $\Omega_r=3\Omega_{\phi}$, as shown in (b). Within one angular period of rotation there are three radial oscillation periods, so the trajectory on the $(y,z)$ plane has a triangulish shape. The dashed lines in (b) show the borders of the phonon spectrum. It can be seen that $\Omega_r$ lies above the optic band and $\Omega_{\phi}$ in the gap of the spectrum. Energies of the rotobreathers are $H=1.03$, 1.11, and 1.27, respectively.}

\textcolor{black}{To support the statements about the dynamic localization of the rotobreather energy, in Fig.~\ref{Fig8new} we present the time evolution of angular coordinates [panels (a)-(c)] and radial coordinates [panels (d)-(g)] for rotobreather excited with $r^*=0.6$ and $\dot{\phi}^*=2.6$. One can see a rapid decrease in vibration amplitudes with distance from the central rotator.}
\begin{figure}[tb]
\begin{center}
\includegraphics[angle=0, width=1.0\linewidth]{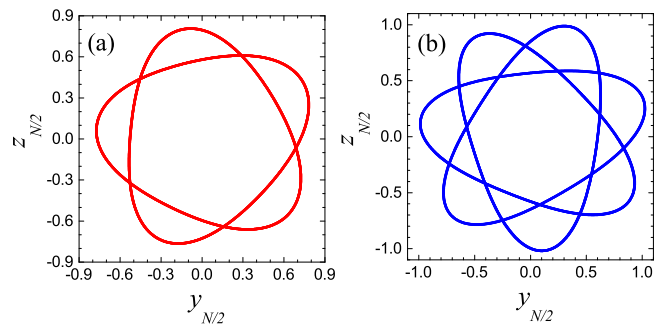}
\end{center}
\caption{\textcolor{black}{Results for regime I. Members of the families of quasiperiodic rotobreathers having closed trajectories with intersections on the $(y,z)$ plane. The rotobreathers were excited with $r^*=0.6$ and (a) $\dot{\phi}^*=3.55$ and (b) $\dot{\phi}^*=5.03$. Model parameters: $r_0=0.5$, $R_0=0.8$, and $k=20$.}
}\label{Fig9}
\end{figure}

\textcolor{black}{For even larger values of $\dot{\phi}^*$ the closed trajectories with intersections on the $(y,z)$ plane can be realized. In Fig.~\ref{Fig9} we present members of the rotobreather families with flower-like trajectories with (a) five and (b) seven petals. The trajectory in (a) closes in two rotations and in (b) in three rotations. The rotobreathers were obtained with the initial parameters $r^*=0.6$ and (a) $\dot{\phi}^*=3.55$ and (b) $\dot{\phi}^*=5.03$. The rotobreather in (a) has frequencies $\Omega_{\phi}=2.55$ and $\Omega_r=(5/2)\Omega_{\phi}$, and energy $H=2.35$. The rotobreather parameters in (b) are $\Omega_{\phi}=2.92$, $\Omega_r=(7/3)\Omega_{\phi}$, and $H=4.63$.}

\textcolor{black}{All the rotobreathers described so far have been obtained for a fixed initial length of the central rotator $r^*=0.6$ and different values of the initial angular velocity $\dot{\phi}^*$, below, close to, and above the angular velocity of vibrationless rotation of single rotator, $\dot{\phi}_v=1.83$.} Our next task is to get rotobreathers with different energies and practically without radial vibrations. \textcolor{black}{For this, different values of the initial length will be considered and for each of them the value of the initial angular velocity will be found at which the rotobreather will have minimal radial oscillations. As a zero approximation for the initial value of $\dot{\phi}^*$ the value $\dot{\phi}_v$ obtained from Eq.~(\ref{dotphi}) is taken. By applying small increments to this estimate, the value $\dot{\phi}^*$ is found that produces the smallest radial vibrations.} 
\begin{figure}[tb]
\begin{center}
\includegraphics[angle=0, width=1.0\linewidth]{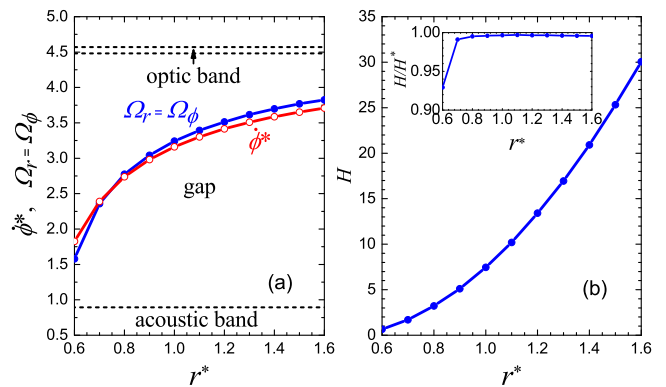}
\end{center}
\caption{\textcolor{black}{Results for regime I. Characteristics of the rotobreathers with minimal radial vibrations as the functions of the initial rotator length $r^*$. (a) Initial angular velocity $\dot{\phi}^*$ (the red line) and equal radial and angular frequencies of the rotobreather $\Omega_r=\Omega_\psi$ (blue line). Dashed lines show the borders of the phonon spectrum. (b) Energy of the rotobreather. The inset in (b) shows the ratio of the rotobreather energy to the energy initially given to the system. Model parameters: $r_0=0.5$, $R_0=0.8$, and $k=20$.}
}\label{Fig10}
\end{figure}

The main characteristics of such rotobreathers are presented in Fig.~\ref{Fig10}. As the functions of initial length of the rotator $r^*$ we plot (a) the initial angular velocity $\dot{\phi}^*$ (red color) and equal radial and angular frequencies of the rotobreather $\Omega_r=\Omega_\psi$ (blue color), as well as the borders of the phonon spectrum (dashed lines); (b) energy of the rotobreather and, in the inset, the ratio of the rotobreather energy to the initial energy given to the system defined by Eqs.~(\ref{Tinit}-\ref{Rn00}). It can be seen in (a) that when $r^*$ decreases approaching the minimum possible value of $r_0=0.5$, the rotobreather angular frequency $\Omega_{\phi}$ decreases rapidly and enters the acoustic phonon band. For this reason we were unable to excite rotobreathers with $r^*<0.57$. \textcolor{black}{Also note that we use sharply localized initial conditions that are not suitable for excitation of less localized rotobreathers with frequencies close to the acoustic band.} When $r^*$ increases, $\Omega_{\phi}$ also increases approaching the optic phonon band. Rotobreathers with angular frequencies above the optical band are impossible, as demonstrated in Sec.~\ref{ElasticRotator} for uncoupled rotators. 
The energy of rotobreathers increases with increasing $r^*$, see (b). \textcolor{black}{The inset in (b) tells us that the energy emitted from the rotobreather increases as $r^*$ decreases and approaches the minimum value $r^*=r_0= 0.5$, which was explained above by the use of sharply localized initial conditions. For $r^*\ge 0.8$, the rotobreather emits less than 0.5\% of the initial energy.}

\textcolor{black}{Looking at Fig.~\ref{Fig10}(a), one could expect a resonance between the second harmonic of the rotobreather and optical phonons. However, the optical band is very narrow, and we did not observe such a resonance due to the relatively large scanning step of the parameter $r^*$.}
\begin{figure}[tb]
\begin{center}
\includegraphics[angle=0, width=1.0\linewidth]{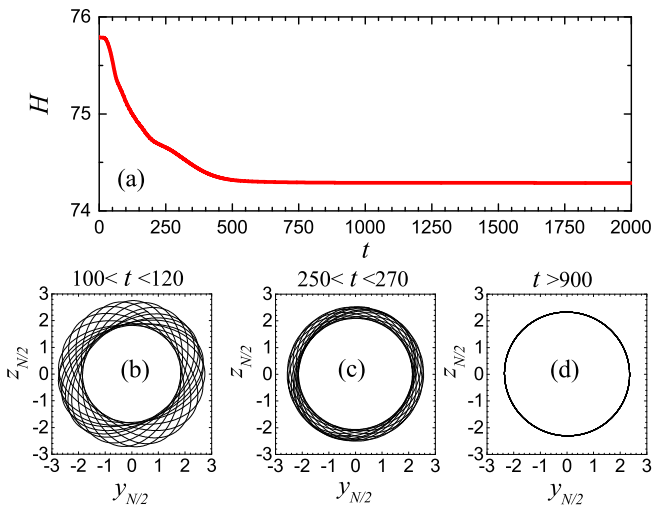}
\end{center}
\caption{\textcolor{black}{Results for regime III. (a) Time evolution of the total energy of the chain with absorbing boundary conditions and central rotator initially excited with $r^*=1.2$ and $\dot{\phi}^*=10$. (b-d) Trajectories of the excited rotator on the $(y,z)$ plane for time intervals specified for each case. In (d), a periodic motion of the rotator is observed, since the trajectory is closed. Model parameters: $r_0=0.5$, $R_0=2\sqrt{2}=2.828$, and $k=20$.}
}\label{Fig11}
\end{figure}
\begin{figure}[tb]
\begin{center}
\includegraphics[angle=0, width=1.0\linewidth]{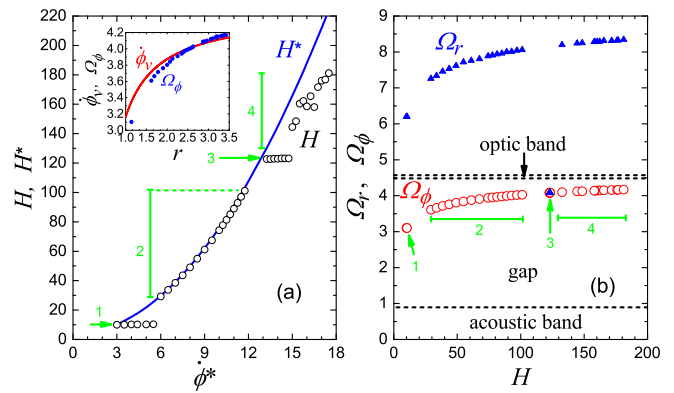}
\end{center}
\caption{\textcolor{black}{Results for regime III. (a) Initial energy given to the system $H^*$ (blue solid line) and energy of the rotobreathers $H$ (open symbols) as the functions of the initial angular velocity of the central rotator $\dot{\phi}^*$ for the fixed initial length of the central rotator $r^*=1.2$. (b) Frequencies of radial and rotational motion, $\Omega_r$ and $\Omega_{\phi}$ as the functions of the rotobreather energy presented by the blue triangles and red circles, respectively. Dashed lines show the borders of the phonon spectrum. The green lines and symbols in (a) and (b) show the energies of the four families of rotobreathers. Model parameters: $r_0=0.5$, $R_0=2\sqrt{2}=2.828$, and $k=20$.}
}\label{Fig12}
\end{figure}

\subsection{Regime III}
\label{RegimeIIIrot}

We assume the model parameters $r_0= 0.5$, $R_0= 2\sqrt 2$, and $k= 20$, for which the gap between the acoustic and optical bands is quite large, as shown in Fig.~\ref{Fig4}(b).

From the condition $dP/dr=0$, where $P(r)$ is defined by Eq.~(\ref{f3}), we find the equilibrium length of rotators $r=0.5974$.

Rotobreathers in the ground state of regime III, described by Eq.~(\ref{groundIII}) with $\phi=-\pi/2$, are excited using the following initial conditions: for the rotator in the middle of the chain, 
\begin{eqnarray}
r_{N/2}(0)=r^*, \quad \dot{r}_{N/2}(0)=0, \nonumber\\ \phi_{N/2}(0)=0, \quad
\dot{\phi}_{N/2}(0)=\dot{\phi}^*, 
\label{IIICone}
\end{eqnarray}
and for $n\neq N/2$,
\begin{eqnarray}
r_n(0)&=&r, \quad \dot{r}_n(0)=0, \nonumber\\ \phi_n(0)&=&(-1)^n\frac{\pi}{2}-\frac{\pi}{2}, \quad
\dot{\phi}_n(0)=0,
\label{IIICall}
\end{eqnarray}
where $N$ is assumed to be an even number.

\textcolor{black}{For the chosen initial parameters $r^*$ and $\dot{\phi}^*$, we observe the dynamics of a system with absorbing boundary conditions. At a sufficiently large initial angular velocity $\dot{\phi}^*$ after a transition period, during which some energy is emitted and absorbed at the boundaries, a rotobreather is formed in the middle of the chain. One example is given in Fig.~\ref{Fig11} for $r^*=1.2$ and $\dot{\phi}^*=10$. In (a), total energy of the chain as the function of time is shown. The energy decreases with time, and after reaching the time $t=900$, the energy reaches an almost constant value, indicating that the radiation of energy by the rotobreather becomes extremely slow.
In (b-d) the trajectories of the excited rotator on the $(y,z)$ plane are shown for time intervals specified for each case. In (d), a periodic motion of the rotator is observed, since the trajectory is closed.}

\textcolor{black}{In regime III, the periodic motion of rotobreathers is always realized for almost circular trajectories in the $(y,z)$ plane, as exemplified in Fig.~\ref{Fig11}(d). In other words, all rotobreathers excited with different initial values of $r^*$ and $\dot{\phi}^*$ reach a periodic regime of motion with minimal radial oscillations, and this is the main difference from regime I, in which several families of periodic rotobreathers with a large amplitude of radial vibrations were observed.}

\textcolor{black}{More information on rotobreathers in regime III can be found in Fig.~\ref{Fig12} obtained for fixed initial length of the central rotator $r^*=1.2$ and different values of the initial angular velocity $\dot{\phi}^*$. In (a), energy $H^*$ given to the chain at $t=0$ is shown by the blue line, while the energy of nearly periodic rotobreathers $H$ is shown by symbols. In (b), radial and angular frequencies of quasiperiodic rotobreathers are shown as the functions of the rotobreather energy by the blue triangles and red circles, respectively. Four groups of rotobreathers can be distinguished, as shown by the green lines and symbols in Fig.~\ref{Fig12}. Excitation with an initial angular velocity in the range $3\le \dot{\phi}^*<6$ produces the same rotobreather 1 with frequencies of angular and radial motion $\Omega_{\phi}=3.10$, $\Omega_r=2\Omega_{\phi}$. An excess of energy given to the chain at $t=0$ is radiated and absorbed at the boundaries. Rotobreathers of group 2 are obtained for $6\le \dot{\phi}^*<11.8$ with very little energy radiation. The radial frequency of group 2 rotobreathers is twice the angular frequency. Initial angular velocity in the range $12\le \dot{\phi}^*<13.3$ does not produces a rotobreather. In this case, no synchronization between angular and radial frequencies is observed, that results in rather strong radiation of energy in the form of radial waves. Initial velocities $13.25\le \dot{\phi}^*<14.75$ produce the same rotobreather 3 with equal frequencies of radial and angular motion, $\Omega_r=\Omega_{\phi}=4.08$. For rotobreathers of group 4 one has $\Omega_r=2\Omega_{\phi}$; they are obtained for $\dot{\phi}^*\ge 15$. Prior to the formation of these rotobreathers a part of the energy given to the chain is radiated and absorbed at the boundaries.} 

From Fig.~\ref{Fig12}(b) it is clear that for growing rotobreather energy its rotational frequency $\Omega_{\phi}$ increases but remains below the optic band, as explained in Sec.~\ref{ElasticRotator}.

\textcolor{black}{Since rotobreathers in regime III perform rather small radial oscillations, it is interesting to see how close are their rotation frequencies to the prediction obtained for single vibration-free rotator, Eq.~(\ref{dotphi}). This information is given in the inset of Fig.~\ref{Fig12}(a). The solid line shows the angular velocity of single vibration-free rotator $\dot{\phi}_v(r)$ and symbols show how the rotobreather angular velocity depends on the radius of its circular orbit. It is seen that the result for single rotator describes reasonably well the circular orbits of rotobreathers.}
\begin{figure}[tb]
\begin{center}
\includegraphics[angle=0, width=1.0\linewidth]{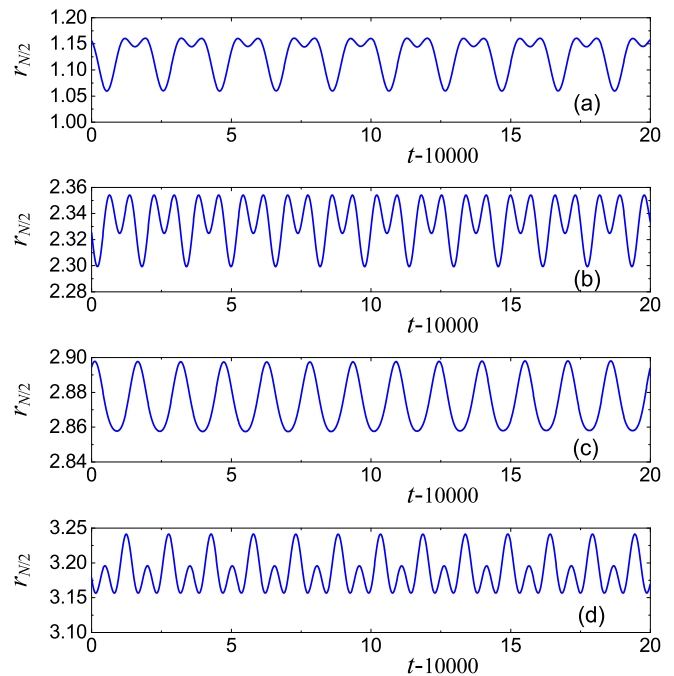}
\end{center}
\caption{\textcolor{black}{Results for regime III. The length of the central rotator as the function of time for quasiperiodic rotobreathers excited with the initial parameters $r^*=1.2$ and different values of $\dot{\phi}^*$: (a) 4, (b) 10, (c) 13.5, and (d) 16. Panels (a) to (d) present rotobreathers of groups 1 to 4, respectively. Model parameters: $r_0=0.5$, $R_0=2\sqrt{2}=2.828$, and $k=20$.}
}\label{Fig13}
\end{figure}

\textcolor{black}{Finally, in Fig.~\ref{Fig13} we give examples of time dependence of radial coordinate of the central rotator for quasiperiodic rotobreathers. Rotobreathers of groups 1 to 4 are are presented in the panels (a) to (d). They were excited with the initial parameters $r^*=1.2$ and different values of $\dot{\phi}^*$: (a) 4, (b) 10, (c) 13.5, and (d) 16. The plots of  Fig.~\ref{Fig13} confirm that the amplitude of radial oscillations in regime III is indeed small, it is 9\% of the averaged rotator length in (a) and does not exceed 3\% in other three cases. Also note that in (a), (c) and (d) $r_{N/2}(t)$ has two maximums per one rotation, while in (b) only one maximum. That is why the frequency of radial vibrations of rotobreathers is twice as high as the frequency of rotational movement in all groups, except for group 3, where the frequencies of vibrations and rotation are equal.}

\textcolor{black}{Rotobreathers do not radiate energy because their radial and rotational frequencies are outside the phonon spectrum of the chain, see Fig.~\ref{Fig12}(b). However, before synchronization between radial and rotational frequencies rotobreather dynamics is described by a number of harmonics and their interaction with phonons results in energy radiation.}

\section{Conclusions}
\label{Conclusions}

Rotobreathers in the chain of coupled rotators with linearly elastic rods and bonds were analyzed numerically. 

Considering dynamics of single rotator (anticontinuum limit), it was shown that the angular frequency of the rotator cannot exceed the frequency of radial oscillations (see Sec.~\ref{ElasticRotator}). 

\textcolor{black}{The chain of rotators can be considered in three different regimes, see Fig.~\ref{Fig2}. Only regimes I and III are considered in this work because regime~II admits different types of ground state structures.}

In Secs.~\ref{RegimeIrot} and \ref{RegimeIIIrot} it was shown that the rotational frequency of rotobreathers \textcolor{black}{in regimes I and III} cannot be higher than the optical band of the phonon spectrum and lies between the optical and acoustic bands [see Figs.~\ref{Fig10}(a) and \ref{Fig12}(b)]. Rotobreathers cannot exist in a chain of rotators with very small $k$, because the gap in the phonon spectrum in this case is either absent [see Fig.~\ref{Fig3}(b) for regime I] or is too narrow, while the second harmonic lies in the optic band [Fig.~\ref{Fig4}(b) for regime III]. Consequently, the conditions for the excitation of rotobreathers improve with an increase in the rigidity of the rods of the rotators, when the gap in the phonon spectrum is large.

These results can be compared with the results given in~\cite{Rotobreathers} for a chain of rotators of a fixed radius (equivalent to the absolutely rigid rods, $k\rightarrow \infty$). The model considered in~\cite{Rotobreathers} allows a rotobreather with an arbitrarily high rotation frequency and does not predict that if the radial stiffness is finite, then the rotobreather frequency will have an upper bound.

\textcolor{black}{In regime III, quasiperiodic rotobreathers can have large-amplitude radial oscillations, as shown in Figs.~\ref{Fig6}, \ref{Fig8} and \ref{Fig9}.} 
Even in such cases, rotobreathers emit energy extremely slowly and have a very long lifetime, since their angular frequency is in the phonon spectrum gap, and the oscillation frequency of the rotator length lies either above the phonon spectrum, or is equal to the angular frequency and therefore lies in the gap. Without resonating with low-amplitude phonons, the rotobreathers do not lose energy for their excitation. 

Radial oscillations of rotobreathers \textcolor{black}{in regime III} can be minimized by proper choice of initial length of the rod $r^*$ and initial angular velocity $\dot{\phi}^*$ of the excited rotator. Parameters of rotobreathers with minimal radial oscillations are presented in Figs.~\ref{Fig7} and \ref{Fig10}.

\textcolor{black}{In regime I, in contrast to regime III, quasiperiodic rotobreathers always have small amplitude of radial oscillations, see Figs.~\ref{Fig11} and \ref{Fig13}.} 

In general, our results describe the dynamic behavior of a chain of elastic rotators. The information presented can be used to qualitatively understand nonlinear dynamics of discrete systems with rotating elastic units, e.g., some polymer chains, etc.

\section*{Acknowledgments}

A.V.S. acknowledges financial support from the Russian Foundation for Basic Research grant No. 18-29-19135 (model formulation, numerical simulations). The work of I.R.S. and S.V.D. was supported by the Russian Science Foundation, grant No. 21-12-00229 (numerical simulations, writing the manuscript).

\end{document}